\documentclass{kapproc} 

\input{psfig}

\let\footnote\savefootnote

\let\footnoterule\savefootnoterule 

\kluwerbib

\def\lessapprox{\setbox0=\hbox{$<$}\setbox1=\hbox{$\sim$}
     \lower0.5\ht0
     \hbox{ \vbox{\baselineskip=0pt\lineskip=0.5pt\box0\box1} }}
\def\greatapprox{\setbox0=\hbox{$>$}\setbox1=\hbox{$\sim$}
     \lower0.5\ht0
     \hbox{ \vbox{\baselineskip=0pt\lineskip=0.5pt\box0\box1} }}
\def\d3k{{\displaystyle {\rm d}{\bf k} \over \displaystyle (2\pi)^3}}

\begin{document}

\articletitle{\centerline{Report to Anaximander:\ }\\
\medskip\\
\centerline{A Dialogue on}\\
\centerline{the Origin of the Cosmos}\\
\centerline{in the Cradle of Western Civilization}}
\chaptitlerunninghead{Report to Anaximander}
\vskip -0.25cm
\author{Rien van de Weygaert}
\affil{Kapteyn Institute, University of Groningen, Groningen, 
the Netherlands}
\email{weygaert@astro.rug.nl}

\vskip 0.5cm
\begin{flushright}
\footnote{Conference summary 2nd Hellenic Cosmology workshop, April 19-20, 
2001}
\vskip -0.5truecm
\end{flushright}
\begin{flushright}
``The Apeiron, from which the elements [are formed], \\
is something that is different''\\
\vskip 0.2cm
Anaximander of Miletus (610-546 B.C.)\\
\end{flushright}                  
                                 
\vskip 0.5cm                
\section*{\rm{\large SCHOOL OF ATHENS ... }}
Pentelic mountain, solid rock in the city of the goddess Athena, patron 
of wisdom. In search of the foundations of our world there 
is hardly a more symbolic site. It welcomed us for a two-day expedition 
to the outer reaches and first instances of our cosmos, a ``symposion'' 
on the origin, evolution and future of the world. 

To Athens Pentelic mountain fulfils a similar role as the city 
itself does to the human quest for the very origins and workings of 
humanity, the world, the cosmos. Its quarries provided the 
``elements'' for the eternal, solemn and awe-inspiring monuments that 
still stand as testimony for an epoch in which humanity reached out 
for unprecedented, thrilling and almost divine heights of intellectual 
endeavour, creativity, and inspiration ! 

The city, obedient to her patron of wisdom, likewise passed on 
the elements and foundation for scientific inquiry. Perhaps unsurpassed in 
beauty and profoundness, it is Raffaello's ``Scuola di Atene'' which 
embodies the most proper expression of gratitude and respect owed by the 
whole world to ancient Greek society. Gratitude for leading the way and bestowing 
upon us the duty to explore and further our understanding of the Universe, for 
showing that one cannot imagine a duty more sacred than this intellectual 
and artistic quest for truth. 

It is therefore with proper modesty that I set out to enumerate 
some of the highlights of these very enjoyable days in April 2001 ... 
days which to us cosmologists stood for a profound and thought-provoking 
experience at the birthplace of science itself ...
\smallskip
\begin{figure}[h]
\centering\mbox{\hskip -0.truecm\psfig{figure=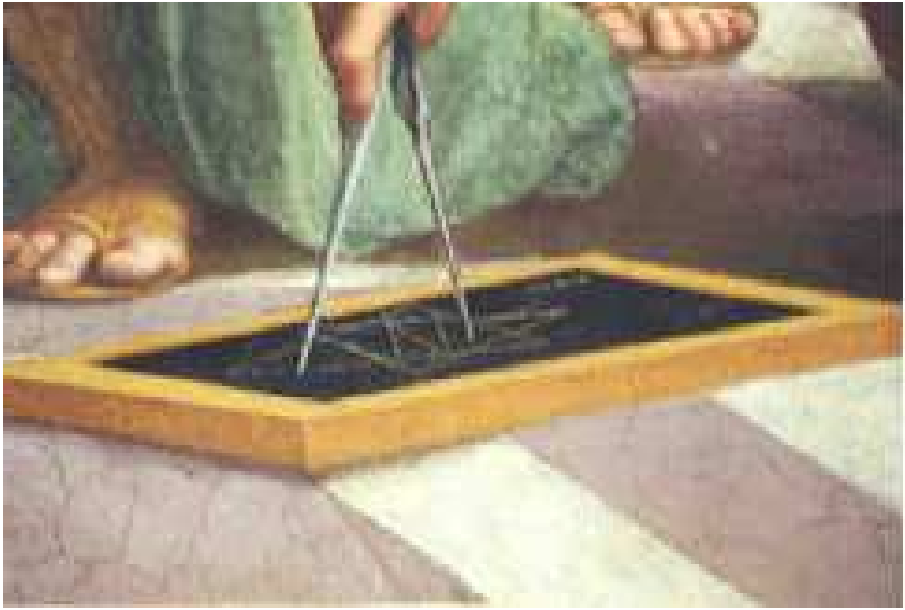,angle=270.,width=12.0cm}}
\vskip -0.25truecm
\caption{Euclides and the geometry of the world.  Detail from 
``Scuola di Atene'' in the Stanza della Segnature, Raffaello Sanzio (1483-1520). 
photo: Rien van de Weygaert 2001}
\vskip -0.5truecm
\end{figure}
\smallskip
\section*{\rm{\large ``THE NOTION OF A POINT IS POINTLESS ... ''}}
As if not yet sufficient, I soon realized that it is futile to strive for 
a summary that evokes the spirit of a meeting to its full and proper 
extent. As was stressed by Spiros Cotsakis, discussing the relativity of matters, 
``the notion of a point is pointless ...''. 

Yet, it were {\it ``lines''} dominating the meeting. {\it ``Wordlines''} first 
and foremost, as cosmologies' principal item. Yet, also the combination of all  
pursued lines of cosmological research whose assembly gave this meeting such 
an tantalizing flair and scope. So varied, nicely complementary and intertwining  
were these lines that the comparison with the construction of a temple of Pentelic marble 
occurred to me as a suitable analogy.

\smallskip
\section*{\rm{\large CONSTRUCTING THE COSMOS ON PENTELIC MARBLE}}
The collection of presentations provided the attendants with a nicely 
balanced and properly representative overview of the beautiful and rich edifice 
into which physical cosmology has matured over the past decades. As it were, 
we were presented with a full plenary gathering of the three branches of 
cosmological architecture. First, there were the sessions of the 
{\bf ``Architects''}. Sessions discussing {\it ``fundamental cosmology''}, 
focussing on the topology and geometry of our Universe and the fundamental 
physical laws and processes at work in those first decisive moments of the 
(Very) Early Universe. Having been provided with these cosmic blueprints, 
the {\bf ``Constructors''} set out to present their investigation into the 
ultimate realization of the cosmic framework. In line with the {\it Classics} of
{cosmology}, they discussed the best available estimates of the parameters 
characterizing the Friedmann-Robertson-Walker metric, by an impressive body 
of evidence the architectural plan for the cosmos of which we have found ourselves 
to be tenants. To them the verdict on the ultimate fate of our 
world ! Finally, the {\bf ``Interior Decorators''} taking care of the 
cosmic {\bf ``Infrastructure''}. It is the realm for the 
cosmological artisans studying the tantalizing and beautiful patterns and objects 
which render our Universe such an awe-inspiring realm of beauty and 
wealth. Part of them kept it clean, concentrating on the works of the 
prime agent responsible for Megaparsec cosmic structure, forming structure 
through gravitational instability. Others were to be seen as the new heroes 
of cosmology, the ones whose {\it ``dirty''} works combine gravitational, hydrodynamic, 
radiative, stellar and a variety of other dissipative processes into a courageous 
attempt towards understanding the lights in the cosmos. In presentations of a 
variety of state-of-the-art projects into {\it ``astrophysical cosmology''} we 
got granted some intriguing pages from the chronicles reporting on the formation 
of galaxies and stars in the Universe, as the lights went on and we got around 
for admiring them ... 

\smallskip
\section*{\rm{\large THE ARCHITECTURE}}
During the late seventies and eighties the image of the ``cosmic snake'' 
emerged as a symbol for a profound link between the physics of the very 
smallest and the physics of the very largest. Cosmology became a forum 
where an exciting mix of views on the workings of our world came to 
meet. Following shortly after overwhelming feats such as the corroboration 
of Big Bang predictions of cosmological light element nucleosynthesis
and the discovery of the Microwave Background Radiation by Penzias 
\& Wilson, the realization had grown that the early Universe provides 
a unique testbed for fundamental physicists and astrophysicists 
alike. It spawned the rise of ``astroparticle'' physics.  The astrophysicists 
yearned for physics providing a compelling set of physical laws.  
Such cosmological theory and framework would allow them to interpret the 
rising tide of observations and understanding of the large-scale 
cosmos. On the other hand, physicists hoped to reduce the ``dirty'' 
Universe of astrophysics into the bare and useful pieces of information on 
the fundamental workings of nature. In the view of some (Veltman 2001) 
this has succeeded in only one instant, astrophysical cosmology having 
taught particle physicists that there are no more than 3 neutrino species, 
constrained by the measured abundances of light elements produced in the 
very first minutes of the Big Bang. 

\smallskip
\begin{figure}[h]
\centering\mbox{\hskip -0.truecm\psfig{figure=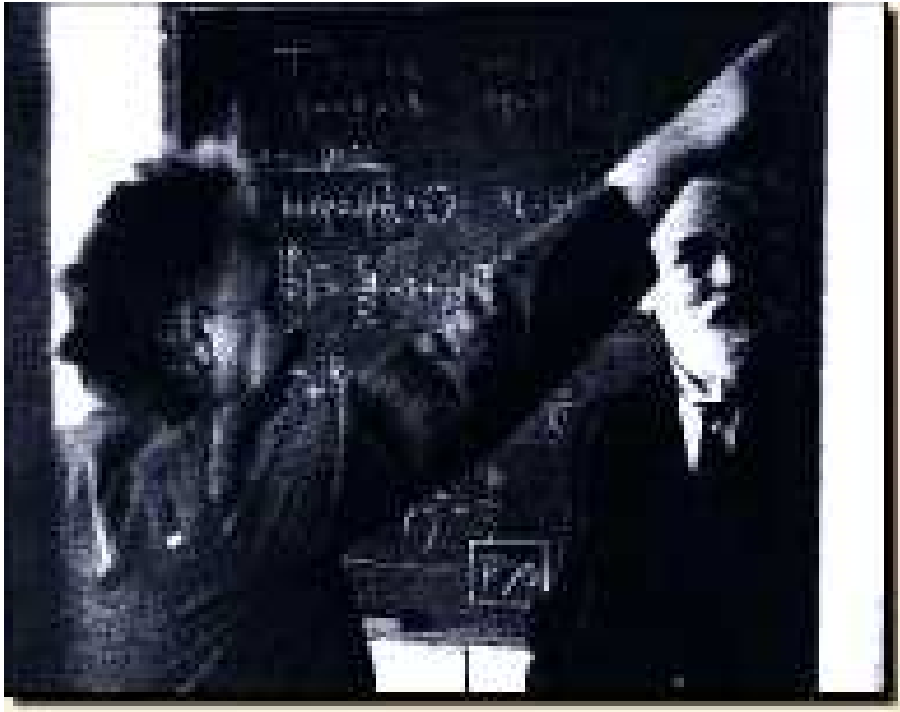,angle=270.,width=12.0cm}}
\vskip -0.25truecm
\caption{Albert Einstein and Willem de Sitter discussing the cosmological equations 
for describing the Universe. Copyright: Wide World Photos, New York. From: exhibit 
Albert Einstein, Center for History of Physics, American Institute of Physics.}
\end{figure}
In this context it is good to realize that one of the most fundamental 
and best understood tenets of cosmology is that General Relativity 
is the proper encryption of the cosmologically dominant force of gravity. That 
while solid thorough experimental evidence -- i.e. other than circumstancial -- for 
GR fully describing gravitational intereactions is still merely confined to the 
puny region of our own Solar System (at least in the view of an orthodox and 
sceptical physicist's view). The Solar System may hardly be considered 
a representative cosmological realm given the fifteen orders of magnitude it 
is removed from the global Hubble radius.  

As much of fundamental importance may be the meaning of a vacuum energy 
density ``cosmological constant'' $\Lambda$.  With some justification we may 
see the evidence accumulated over the past years for a non-zero value 
of $\Lambda$ as the currently most significant and fundamental development in the 
field of cosmology. It may provide us with a key for unlocking 
the workings of gravity at the very first instances. The quest for the origin of 
the cosmological constant and its possible significance as a fundamental trace 
of quantum cosmological processes is, in my view, one of the most exciting 
and illustrious scientific endeavours in early 21st century physics. It brings 
us to the very first moments of our Cosmos, an epoch at which it resided in a 
still largely mysterious era of quantum gravity. It may be here that we can 
expect to run into the very secrets of the ultimate creation of the Universe. 
In the light of such considerations there is some urge in investigating the 
foundations and assumptions upon which our ``standard'' 
cosmological framework stands. A more lucid understanding 
of the specific position of the standard Friedmann-Robertson-Walker 
within a wider context of cosmological theories is necessary. 
Various contributions involved attempts to adress the issue of this  
special role and position of FRW Universes. The string of presentations 
on fundamental cosmological issues in a general relativistic context 
constituted a fruitful and useful forum for further contemplations along 
this line, triggering many stimulating thoughts.  

For natural sciences, it is ultimately compelling to turn to the 
mathematical framework in which the science is enabled to mature. Only 
then the full richness of links and parallels can be discerned which are 
so necessary for uncovering the fundamental issues and laws. 
Therefore, it was the contribution by S. Cotsakis which functioned 
as a beautiful frame for the works on the issues of fundamental 
cosmology. To see cosmological world models treated as mathematical objects 
is particularly usuful to the astrophysical community, making them 
to realize into which narrow lanes they tend to channel their results and 
observations, ad well as for grasping the wider significance of their 
work. Outlining the 3 basic ingredients which constitute a complete theory -- the 
corresponding theory of gravity, the spacetime structure and geometry, and the matter 
fields which form its content -- it paved the way for contributions on 
specific elaborations on various specific topics. 

Directly tangible is the ingredient of matter and energy content of 
our Universe. Still puzzling are the nature of dark matter, so tangible 
present in its gravitational influence, as well as the elusive vacuum energy 
which is responsible for the cosmological constant $\Lambda$, which in recent 
years seems to have surfaced as an inescapable presence. As for the nature of 
dark matter, its properties can best be determined from its presence on 
galactic scales, in particular its immediate influence on the rotation 
curves of large galaxies. Justifiable a discussion on this issue was 
a necessary ingredient within this meeting, taken care of in the contribution 
by Spyrou. Also with respect to the untangible component of vacuum density,
the current popularity and high interest in the cosmological constant did not leave 
the capital of Attica untouched. It was entertaining to notice how the symbol of 
its traditional rival in Laconia, $\Lambda$, still managed to incite 
hot discussion and contention ... for sure still a sensitivity that does not 
seem to have subsided ...

The second ingredient, that of spacetime geometry, got further elaborated on by Christodoulakis 
and G. Papadopoulos. They discussed the specific spacetime structure of a few classes of 
spatially homogeneous Bianchi cosmologies. Centering more specifically   
on the widely accepted Friedman-Robertson-Walker Universes, Tsamparlis 
tried to answer the question whether or not these models do also form the 
worldmodel in situations where symmetry assumptions are less strong. Intentionally 
focussing on the issue of time, Kehagias traced and investigated its extreme 
ramifications in the context of time machines. 

Combining all three basic ingredients, including those of the gravitational 
force, Cotsakis' extensively discussed the stability considerations of such 
mathematical world models. Particularly stressed was the fundamental fundamental significance of 
{\it singularities} (``functions, just like living beings, are 
characterized by their singularieis''). Drawing particular to the global FRW Universes, 
Cotsakis finished this issue with a discussion of the important question of the stability of 
the global FRW Universes. Whether it is indeed a stable solution along the lines of 
cosmological attractor theory is highly relevant for simple perturbation considerations in 
conventional structure formation considerations, which tend to neglect the fact that these 
are less well-posed in a full relativistic setting than often presumed. 

Given the statement that nearly all viable cosmological solutions are likely to form 
singularities in a finite time, we quickly arrive at the overriding question whether {\it 
quantum gravity} is really necessary to resolve these singularities. If so, implying 
the need for a bigger encompassing theory of cosmology, we may indeed get to 
the inescapable conclusion that the notion of a point is rendered obsolete, for 
sure in those very first decisive instances. Elaborating on this issue of a 
fundamental connection between the quantum world and gravity, it may be found through 
the elusive $\Lambda$ term. Elizalde focussed on this connection by discussing the Casimir 
effect, which should help us in understanding the gravitational action of 
the vacuum energy density.

Extrapolating the discussion on the cosmological constant $\Lambda$ to the very first 
moments of the Universe, its present value may be a mere small remnant of the 
overriding power it unfolded during early phases of inflationary expansion. This 
provided the backdrop for the discussions by Papantonopoulos, Miritzis and Dimopoulos. 
Embedding inflation within a multi-dimensional brane cosmology is one of the currently 
investigated attempts to develop a proper understanding of why we ended up in a 
FRW Universe (Papantonopoulos). Very interesting would be the possibility of 
inflation leaving an observational and measurable imprint in the form of primordial 
magnetic fields, the seeds for the currently not yet understood relatively strong 
galactic magnetic fields. The origin of these galactic fields is still 
an as yet unsolved riddle in the astrophysical realm. The suggestion that they 
may have been a remnant of cosmic inflation provides an intriguing and relevant 
connection between the world of the very early Universe and that of our 
observable Universe. Tsagas subsequently discussed the possibility of such 
cosmic magnetic fields so strongly coupling to the spacetime geometry that they 
would not leave the universal expansion untouched. 

\section*{\rm{\large THE CONSTRUCTION}}
The architects finally agreed on handing over a world model scheme to the 
constructors, in the form of the homogeneous, isotropic and uniformly 
expanding Friedmann-Robertson-Walker model Universe. This agreed upon, discussions opened 
on the precise parameter values which characterize its realization in the 
case of the world we live in. While great strides forward have been made with respect 
to setting the values the crucial cosmological parameters -- the curvature of 
spacetime, the mass-energy density $\Omega$ in terms of the critical density 
and the expansion rate in terms of the Hubble parameter $H_{\circ}$ -- not all 
have been determined with comparable accuracy. 

An extensive and very informative presentation by Gazta{\~n}aga centered  
on the large scale distribution of matter and galaxies and the microwave 
background radiation as seemingly inexhaustible sources of information on 
the fundamental cosmological parameters. It set the scene for the contributions 
on the first day of the meeting, providing the background for all ensuing 
presentations on structure and galaxy formation. 
\begin{figure}[b]
\vskip -0.5truecm
\centering\mbox{\hskip -1.0truecm\psfig{figure=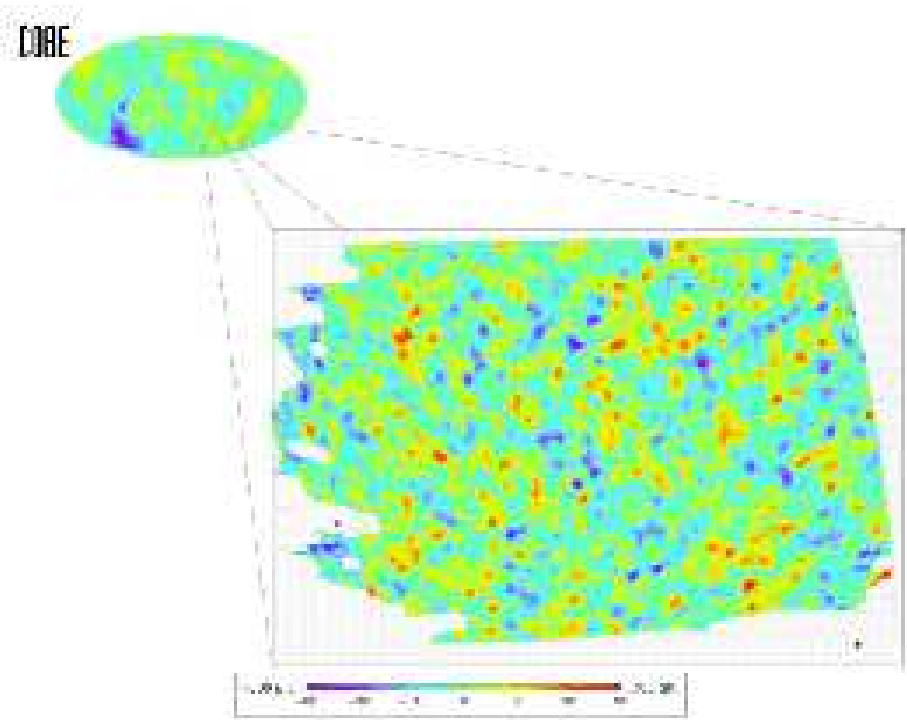,angle=270.,width=13.5cm}}
\vskip -0.25truecm
\caption{The Cosmic Blueprint (?): 
Boomerang map of the Microwave Background, image obtained from the maiden 
voyage of the balloon-borne instrument around the Antarctic (2000). The BOOMERANG image 
covers approximately 2.5
revealing hundreds of complex structures that are visible as tiny variations -- 
typically only 100 $\mu$K -- in the temperature of the CMB. The patterns 
visible in the image are consistent with those that would result from sound waves 
racing through the early universe, creating the structures that by now have evolved 
into giant clusters and super-clusters of galaxies. Courtesy: Boomerang Collaboration.}  
\vskip -0.5truecm
\end{figure}
During many decades of cosmological observations, the overall construction plan  
in terms of the FRW blueprint has accumulated an impressive record of observational 
evidence. Of decisive influence has been the discovery of the perfectly thermal 
microwave background radiation field. The thermal spectrum of the microwave background, 
outlining the surface of last scattering at the epoch of recombination, has been 
measured by COBE's FIRAS instrument to a degree of precision inconceivable to the 
vast majority of astronomical observations. Its tight Planckian blackbody behaviour 
is the most convincing evidence for the reality of the early hot and highly 
dense epoch (the `Hot' of the Big Bang) out of which the present Universe emerged 

Despite such solid and practically irrefutable evidence, the basic specification 
of the FRW Universe description in terms of its curvature, the expansion rate, and 
the content in matter and energy is still not settled to any desirable and 
needed precision. However, we have witnessed great progress during the past 
two decades, and cosmology has managed to break away from a bare search for two numbers and 
develop into one of sciences' most prospering activities, encompassing and 
combining a rich and sumptuous palet of relevant areas of interest. Even more 
impressive, possibly conclusive, advances are shortly to be expected. It therefore 
does not anymore seem an exaggeration to claim that the fundamental cosmic parameters 
will get fully fixed within the near future. Within a decade we may have 
fully determined the past, history and fate of our Universe. Nonetheless, if blessed 
with some feeling for historic awareness we should be rather weary of such firm 
statements and claims. After all, with a Universe whose energy content is seemingly 
dominated by a dark matter component and an additionally even more enigmatic dark energy 
contribution responsible for the cosmological constant $\Lambda$ it may not be entirely 
unwarranted to suspect that we are missing out on some fundamental and systematic aspect 
of our worldview. After all, the physical nature of the dark matter is currently still as 
puzzling as it was more than a decade ago. Even more humbling we should be with 
respect to $\Lambda$, the essence of the corresponding vacuum energy density 
even being further beyond our grasp (see discussion above, and Weinberg, S., 
1989, Revs.~Mod.~Phys. 61, 1).

The value of the curvature parameter $k$ has been converging strongly towards 
that of a flat $k=0$ value. The work involved with the high-redshift Supernovae SNI 
has been yielding substantial support towards this geometry. Almost inescapable 
evidence has been provided by the Cosmic Microwave Background balloonborne experiments. 
The MWB balloon experiments -- currently most prominently represented by the 
Boomerang and Maxima project(s) -- have provided us with an unprecedented, detailed 
and tantalizing view of the pristine Universe (Fig. 3). Their measurements so strongly suggest  
a flat $k=0$ Universe that to escape this value would almost imply a flaw in our 
understanding of MWB physics. With only a few years since the detection of the first 
Sachs-Wolfe fluctuations on an angular scale of $\theta \approx 7^{\circ}$ by the 
COBE satellite in 1992, one cannot help but being impressed by the achievements 
of the balloon measurements at a vastly more detailed resolution of $10-20$ arcmin. 
With the newest results having an effective resolution allowing 
the detection of even a second and third Doppler peak in its angular fluctuation 
power spectrum, it is hard to escape the conclusion that the Universe 
is indeed a flat one, $k=0$.

More problems still seem to be involved with the value of the current 
expansion rate of the Universe, encapsulated in terms of 
the Hubble parameter $H_{\circ}$. Progress has been substantially hampered  
by the lack of really significant advances in the quality of distance estimators, with 
which we mean improvements by an order of magnitude. Yet, in comparison to 2 decades ago, 
an enormous investment of effort has been yielding a gradually convergence 
of estimates of $H_{\circ}$ onto a tighter and tigher region of acceptable 
values. Indeed, the main HST Hubble project has been succesfull in the sense 
of having produced convincing constraints on a value of $H_{\circ} \approx 70\ 
\hbox{\rm km/s/Mpc}$. 

Leaves us the issue of whether we are really closing in on settling the cosmological energy 
density content. The radiation contribution $\Omega_{radiation}$ is known 
to high precision, ever since COBE demonstrated the precision of the Planck spectrum 
and nailed its temperature to $T=2.728\pm0.004\hbox{\rm K}$ we may consider this issue 
to be set. Once we have established the value of the Hubble parameter we then have 
a firm estimate for $\Omega_b$, the contribution by plain baryons to the cosmic density. Its 
value of $\Omega_b\approx 0.043 \pm 0.03h^{-2}_{65}$ is tightly fixed by the 
light element abundances produced during the cosmological nucleosynthesis processes 
in the very first minutes of the Big Bang. Telling for the rapid progress in this 
field is the fact that at the meeting the seemingly low amplitude of the Doppler peak 
in the first round of Boomerang results were still discussed in the sense of being 
problematic for the value of the cosmic baryonic matter density. In the meantime 
new Boomerang results have branded this problem as a non-existent one. Leaves 
us to mention the estimated contributions of matter in total and the contribution 
involved with $\Lambda$. The currently favorite value for the total matter density 
is $\Omega_{matter} \approx 0.3$, including the enigmatic dark matter 
contribution. However, one cannot help but sensing that a clear view of matters 
still seems to be removed beyond our immediate grasp. Interestingly, the 
stated $\Omega_{matter}$ value and the estimate for the vacuum energy density, 
$\Omega_{\Lambda} \approx 0.7$, have acquired a status of ``standard'' model. 
However, one cannot be completely unjustified in feeling rather uneasy about the 
wide agreement on values whose inherent uncertainty should suggest one to 
expect a large spread in measured values. Purely because of arguments based on 
simple freshmen statistics !

Despite some side remarks on present-day affairs we may go along with the optimism 
of Gazta{\~n}aga's contribution on the final settling of the Universe's parameters. It 
will be the MWB experiments that will fully settle the values of some 15-18 
fundamental cosmological parameters at presently unimaginable levels of accuracy 
at sub-percent level. It will be up to the recently launched MAP 
satellite and ultimately the superior performance of the European Planck satellite 
to settle the record.  

It is therefore not entirely unreasonable to be delighted with respect to upcoming 
events, enthusiasm which possibly may be tempered when considering your subsequent 
job prospects as cosmologist. To some extent these experiments may signify the end 
of cosmology as the science involved with measuring the parameters specifying the 
structure of the global Universe. Roads of cosmologists may split into two different 
directions. One road will lead back into time, towards the very first instances in which the 
many truely fundamental riddles await us for further elucidation. Others may 
be inclined to follow the Universe's direction of time, and pursue their study 
forward from the post-recombination Universe onward. 

Naturally, the global cosmological parameters are not only affecting the 
evolution of the Universe on the very largest scales of the Hubble radius. 
Gazt{\~n}aga assessed the way in which the cosmological parameters 
will be reflected in the kinematic and structural evolution on more local  
cosmic scales. The morphology, kinematics and dynamics of the Megaparsec 
cosmic foam, of clusters, and even of galaxies will be influenced to 
a considerable extent by the properties of the global Universe in which 
they emerge. In turn, understanding these influences provides us with 
additional means of determining the cosmological parameters from 
local measurements. This is currently also one of the most pursued 
strategies, profiting from the large amount of available and well understood 
observational information. 

In the end local measurements may yield cleaner signals than feasible on 
the basis of the Supernovae SNI results. The latter still pose us with questions 
concerning the intricate, complex and only partially understood physical 
processes in the supernovae. Equally interesting will be to compare these 
local measurements with those yielded by global determinations, it may inform 
us of systematic problems with our world models. One example of such 
situation would be a cosmological evolution of fundamental physical 
constants, as some have claimed the fine structure constant $\alpha$ is 
doing. In line with such considerations, Perivolaropoulos dicussed the possibility 
of identifying the effect of a finite cosmological constant on the dynamics of 
clusters, groups of galaxies as well as those of galaxies. His conclusion 
that clusters of galaxes may indeed evidence of detectable dynamical effects 
of a cosmological constant (while the available technology will not be able to 
find any comparable clear signature on the scales of galaxies) again stresses  
the significance of the dynamically young yet outstanding clusters of galaxies 
as important cosmological laboratories. 
\smallskip
\begin{figure}[b]
\vskip -0.5truecm
\centering\mbox{\hskip -0.truecm\psfig{figure=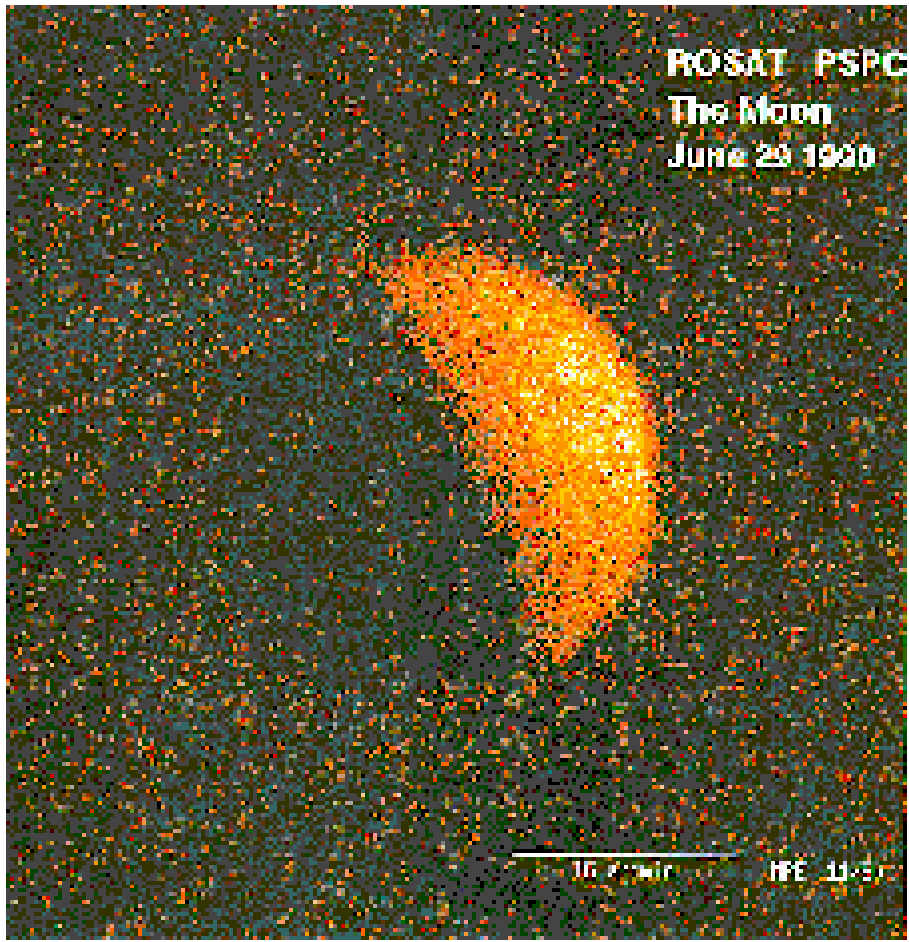,width=12.0cm}}
\vskip -0.25truecm
\caption{The X-ray cosmic background exposed ! X-ray image of the moon in the 
0.1-2 kEV. Note the decrease in the XRB in the dark side of the moon 
(from Schmitt et al., 1991, Nature 349, 5583). Courtesy: MPE, Garching, Germany.}
\vskip -0.5truecm
\end{figure}
\smallskip
\section*{\rm{\large THE INFRASTRUCTURE}}
Gazta{\~naga} henceforth laid out the context of the subsequent contributions by 
describing how from the parameter specification of the cosmic blueprint we can 
set out to study the development of its internal organization. Once the global 
cosmic parameters have been fixed and set, we may establish a proper and clean 
basis of initial cosmological conditions. 

One of the major developments in our cosmological worldview over the past 
quarter of the century involved our radically changed view of the 
spatial organization of matter in the Universe. Not only does it involve an 
interesting issue in its own right, and does it involve the issue of how 
the galaxies and stars have finally emerged. It may indeed be claimed to hold 
the key towards solving the very riddle of the Universe itself. Its origin can 
be traced back to the linear embryonic structure at the epoch of 
recombination. Its properties, via the specifying power spectrum, holds the 
most direct observable link to the physical processes operating at the 
very first stages of our universe. Ultimately, we may therefore even hope to use 
the accumulated understanding of the cosmic structure formation process 
towards solving the thrilling riddles of the very nature and origin of the 
Cosmos itself. 

To study the emergence and evolution of structure within the Universe we 
need to identify those structures and natural processes and phenomena that 
still have retained a memory of earlier cosmological conditions. Within 
our cosmos the Microwave Background (Gazta{\~n}aga), Megaparsec scale cosmic 
structures (Basilakos, Georgantopoulos, Plionis, van de Weygaert) and even 
the rotation of galaxies (Voglis) are examples of such precious 
{\it ``COSMIC FOSSILS''}. These cosmic fossils will enable us to reconstruct 
the history of the Universe. The cosmic history from the Recombination Epoch 
onward, and perhaps even that of the preceding eons !!! 

The Holy Grail of many structure formation studies is the recovery of the 
fluctuation power spectrum. Once we have a situation in which the primordial 
density and velocity perturbations are 
of a basic Gaussian nature and in which the gravitational instability mechanism 
functions as the prime mover of the migration currents of matter through their 
cosmic volume the power spectrum fully specifies the evolution of the internal 
matter distribution. As for the two conditions, it has yet again been the 
COBE satellite which produced the most substantial and solid evidence for 
the reality of these circumstances. Apart from a few recent claims for minor 
(measured) deviations, the basic Gaussian nature of the patterns on the surface 
of last scattering is irrefutable. As for the gravitational instability mechanims, 
COBE was not only an overwhelming success for discovering the reality of the 
fluctuations in the pristine Universe. Equally important almost was its 
finding that the amplitude of the fluctuations was almost fully in accordance 
with the expected values implied by the density contrast of structures in the 
local, present-day, Universe. It may indeed be considered as one of the most 
impressive triumphs of cosmology that it managed to tie together 
structures over a vast range of spatial scales and predict on the basis 
of the structures on Megaparsec scale how large the Sachs-Wolfe 
fluctuations in the MWB would approximately be (except for some minor 
modifications). 

Once we have mapped out the primordial field of density and velocity 
perturbations on the surface of last scattering (given the global cosmic 
context and the relative contributions of baryonic matter, dark matter and 
radiation to the energy density of the Universe) we have at our disposal 
all necessary information to set out a detailed track for the evolution 
and emergence of structure in the Universe. We may then finally hope to 
develop a properly detailed understanding of the formation of the cosmic 
matter distribution into its salient Megaparsec patterns, and 
even obtain a reasonably detailed view of the way in which 
galaxies have formed and the stars started to shine. When indeed we 
have reached this stage, astrophysical cosmology will be able to 
focus exclusively on the purely astrophysical phenomena and processes 
that shaped the infrastructure of the Universe. 

For the determination of the form and amplitude of the primordial power 
spectrum which underlies the formation of all structures and objects 
in our Universe, we need to combine information from a large variety of 
sources, structures and physical processes. With each structure, process and 
object corresponds a characteristics spatial range at which they 
are optimally sensitive to the original power spectrum. 

The MWB experiments provide the most direct and cleanest power spectrum signal  
available. In principle the CMB yields information on fluctuations ranging from 
the very largest scales (the COBE resolution of $\theta\approx7^{\circ}$ corresponds 
to structures in the order of a Gpc) down to scales comparable to supercluster and 
cluster scales. The clustering of conspicuous and outstandin objects like AGNs and 
quasars can provide useful information on intermediate scale fluctuations. However,  
we still do not really understand their role in the cosmic scheme of structure formation 
(see contribution Basilakos). In the end, this is a necessary condition if we wish to 
relate their clustering to the underlying matter distribution. 

On scales of a hundred to a few hundred Megaparsec we may hope to infer 
some useful information from the peculiar motions of galaxies. However, uncertainties 
in distance estimates still poses overriding problems for such estimates to 
improve in quality in the foreseeable future. Supreme signatures are already being 
inferred on the basis the weak lensing signatures by large scale structures on similar 
scales. These measurements are indiscriminately sensitive to the full matter content and 
hence ``dark'' matter proof. This makes them a source of tremendous future potential and 
promise, a promise that materialize soon now sensitive wide-field instruments 
are rapidly being commissioned into operation. The ``cosmic shear'' 
in deformed background galaxy images has already been translated 
into a beautiful set of measurements of $P(k)$ over a range of quasi-linear 
and linear multi-Megaparsec scales. 
 
At the small scale end, down to a scale in the order of a Megaparsec, we are 
confined to the distribution of galaxies in the cosmic web (see contributions 
van de Weygaert and Plionis). It is here that 
careful statistical analysis of uniform galaxy samples have played an instrumental 
role. Gazta{\~n}aga sketched his work on the APM catalogue of galaxies, as yet still 
the largest sample of (projected sky) galaxy locations available. We may hope 
that with the completion of the SDSS redshift survey, as well as from the accompanying 
deep five-colour photometric sky survey, we will be able to extract statistically 
equally clear results on the full 3-D galaxy distribution. There have been some 
tantalizing indications from the recently completed 2dF survey for an interesting 
feature at the interesting scale of around $100h^{-1}\hbox{Mpc}$, a feature 
that has also been claimed to be observable in the APM power spectrum and as well 
in the new CMB Boomerang and Maxima data at $k\approx 0.06-0.6$. It would indeed be 
a thrilling and highly meaningful result, suggested by some to be the result from 
a phase transition associated with spontaneous symmetry breaking in the early Universe. 
Nonetheless, the reality of the signal is still contrived, and it may have 
surfaced as it occurs on the fringe of what is technically feasible.

At much smaller scales direct measurements of the primordial power spectrum 
will become increasingly difficult to obtain and to isolate from other effects. 
The formation and shaping of structure gets increasingly influenced by the full 
arsenal of dissipative hydrodynamic, radiative and stellar processes. Therefore, 
it would be unjustifiably optimistic to claim that sufficiently unaffected 
information on the (primordial) power spectrum $P(k)$ can be extracted 
on these subgalactic scales. Nonetheless, theoretical and numerical work has 
helped to show that even on such nonlinear scales we may find direct traces to 
primordial circumstances. In fact, some of the most conspicuous and telling 
traces in our astronomical world (see Voglis). 

\subsection*{{\it Infrastructure: \hskip 0.5cm Gravity and the Web}}
Over the past two decades we have witnessed a paradigm shift in our 
perception of the Megaparsec scale structure in the Universe. As increasingly 
elaborate galaxy redshift surveys charted ever larger regions in the nearby 
cosmos, an intriguingly complex and salient foamlike network in the cosmic matter 
and galaxy distribution got to unfold itself. This distinctive foamy pattern is 
characterizied by galaxies accumulating in walls, filaments and dense compact 
clusters surrounding large near-empty void regions (see contribution van 
de Weygaert). As borne out by a large sequence of computer experiments, such 
weblike patterns in the overall cosmic matter distribution do represent 
a universal but possibly transient phase in the gravitationally propelled 
emergence and evolution of cosmic structure. 

The gravitational dominance in the shaping of cosmic structure is mostly 
confined to that of the global Universe down to scales of a few 
Megaparsec. The cosmic large scale structure and the various elements of the 
cosmic foam have been such a prominent item of scientific interest over the 
past few decades because of this direct relation to the circumstances 
in the Early Universe, the ideal ``cosmic fossil''. 

On galaxy scale, however, there is at least one prominent aspect 
which relates straightly to the gravitational formation process itself. In 
hierarchical scenarios of structure formation the rotation of galaxies is 
induced by the gravitational torqueing of the collapsing density peaks. The 
tidal torque generation of galaxy angular momentum was the subject of the 
presentation by Voglis. An interesting observation has been the finding 
of galaxy cores that counterrotate with respect to the main body of 
the galaxy. Voglis discussed the possibilities of explaining and working 
out such a phenomenon within scenarios where a core would result from 
a massive merger event or where it would result from a successive infall 
from many small clumps. It resulted into a solid, mathematically well-founded, 
expose on important dynamical aspects. Indeed, it opened insights into seeing 
how the existence of such cores may be linked to the cosmological conditions, 
which makes it into a welcome small-scale ``cosmic fossil''. 

Moving up in scale, we know that the Megaparsec cosmic web is most 
characteristically outlined by its filamentary features. These filaments result 
from the anisotropic gravitational collapse of overdense matter regions. With 
this in mind Plionis set out to describe how one can exploit their 
properties to understand the dynamical development of clusters and 
filaments in hierarchical structure formation scenarios. Filamentary patterns 
correspond to intermediate evolutionary phases, having established a 
pronounced appearance and contrast with respect to the surrounding 
Universe but not yet having fully contracted onto self-gravitating 
and virialized entities. Within the context of the cosmic foam in which 
they are embedded, these anisotropic structures form connecting bridges 
between its most compact elements, the clusters of galaxies. The evolution 
of cosmic structure involves a systematic migration of matter towards 
the high-density regions, in which filaments function as 
channels along which matter gets gradually transported towards clusters. 
Given their dynamical youth within the cosmic sheme, clusters being the 
youngest fully collapsed components and filaments still in an intermediate 
yet pronounced stage, they should form one of the most 
dynamically active sites in cosmic Megaparsec scale structure. 
These Megaparsec scale structures are the present-day equivalents of 
the similarly active but much smaller regions at earlier cosmic epochs, at 
an epoch at which the cosmic matter distribution resided in a dynamically 
younger phase. A closer assessment of the processes involved with these 
filament and cluster junctions is compelling in that it will inform 
us about the very buildup of the cosmic web structure itself. 
Plionis therefore focussed specifically on the ramifications of the implied 
anisotropic inflow of material along filamanets onto clusters. On the basis 
of a sample of 952 APM clusters he presented convincing evidence for alignments 
between neighbouring clusters. Moreover, the dynamical youth of the clusters is 
underlined by the presence of significant substructure in some $30-40\%$ 
of the clusters. Most strongly and telling was the finding of the 
alignment of these substructures with the surroundings, highly suggestive 
of their inflow along the anisotropically outlined channels formed by the 
filaments.  

While Plionis focussed on the aspect of the filamentary elements of the 
cosmic foam, using the presence of relevant and readily exploitable 
observational material to study its properties in the real universe, Van 
de Weygaert choose to emphasize the geometric nature of the cosmic foam 
in an attempt to provide a universal theoretical frame for its distinct 
nontrivial morphology. He described the intrinsically mathematical 
concept from the field of ``stochastic geometry'', the Voronoi model. 
The geometric Voronoi model seeks to exploit the distinct geometric 
nature of the patterns in the cosmic matter distribution. It attempts 
to combine two crucial and salient properties of the cosmic foam. 
Firstly, the stochastic (yet, not Poissonian !!!!) spatial setting of 
distinct geometrical elements, walls, filaments and high density 
nodes (clusters). Secondly, the distinct cellular morphology which forces 
the elements of walls, filaments and vertices in a definite and tight 
mutual relationship to each other. Clusters are found at specific locations 
(vertices) within such geometric networks, at the interstices where 
filaments and walls connect. In turn, the filaments and walls surround  
and thus avoid the large voids, which take up most of the spatial volume in 
the Universe. On the basis of these properties, which are indeed prevalent 
in the real Universe as well as in N-body computer models of gravitationally 
induced structure formation, Van de Weygaert showed the presence of a 
definite ``geometric bias'' in the superclustering of clusters. The most 
massive clusters will, as a consequence of their distinct spatial locations, 
form massive supercomplexes, whose sizes may supersede the basic 
voidsize by an order of magnitude. Geometric simulations revealed the 
presence of seemingly flattened superclusters over enormous spatial 
scales of hundreds of Megaparsec. Mathematically, it seems to indicate 
a beautiful, hitherto unknown, pattern of similarity. Cosmologically, it 
may form an indication for the presence of a purely ``geometric'' biasing 
effect, a ramification of the observed foamlike galaxy distribution. 

Indeed, the clustering of different populations of special astrophysical 
objects may be easier understood within the context of taking into account 
the context of the cellular geometry of the matter and galaxy distribution 
in the Universe. An important aspect of this was touched upon in the 
contribution of Basilakos. Over the years, there have been various attempts 
to measure clustering in the Universe on the largest scales, far exceeding 
the clustering of normal galaxies on scales of a few up to a few tens of 
Megaparsec. Both clusters of galaxies as well as classes of Active Galaxies, 
quasars and radio galaxies are noteworthy examples. The high luminosity of 
AGNs makes them ideal to probe cosmic structure over vast reaches of the 
observable Universe. As they form a sparsely sampled reflection of the 
underlying patterns and appear to be located in or near the highest density 
regions within the cosmic network, they are a valuable complementary 
population for tracing cosmic structure on the largest ($> 100h^{-1}\hbox{Mpc}$). 
Conventional approaches were often hampered by the ill-defined selection 
criteria of the AGN samples. Proper selection is of utmost importance, even 
more so as the physical nature and therefore location of these sources within the 
cosmic matter distribution still evades full understanding. Basilakos managed to get 
rid of a few of such considerations, by studying cleanly defined samples on the basis 
of uniform X-ray selected samples of AGNs. This allowed him to adress in 
a clearly defined fashion the question of the particular ``bias'' of such 
source populations with respect to the large scale structure outlined 
by normal galaxies. Within a set of structure formation scenarios 
(mainly $\Lambda CDM$ and $\tau CDM$), three different bias descriptions 
were tested. These ranged from bias-free to the population evolving 
merging bias prescription of Mo \& White. While it allowed to put constraints 
on the validity of the scenarios, and corresponding bias descriptions, the 
main theme was the wider scenario-independent conclusion that indeed 
AGNs can offer us a valuable tracing of structure on the very largest 
scales otherwise unreachable by normal galaxy samples. 

With Basilakos having stressed the importance of AGNS for studying large 
scale structure, and having argued how their X-ray emission can be exploited
as an optimal selection criterion, it is a natural step to extrapolate this to 
the high redshift Universe. 
We know that at high redshift the volume density of such X-ray emitting 
AGNs and quasars is far higher than locally (peaking at around $z \approx 1-2.5$), 
and it may therefore not be a surprise that the first X-ray satellite, before 
the discovery of the MWB background, discovered an almost isotropic background 
of X-ray radiation (to within a few percent), except for a dipole component completely 
in accordance with that in the MWB. Still, I consider the wonderful ROSAT picture of 
the obscured moon as one of astronomy's ``richest'' symbolic images (Fig. 4). Georgantopoulos 
described the exciting recent developments in unravelling the secrets of 
this diffuse X-ray background. Since the operation of ROSAT and in particular since 
the superb spatial and spectral resolution of the new Chandra and XMM X-ray satellite 
observatories have been put into operation, the essential 
significance of the X-ray background may be regarded as settled. For many 
years, it was not clear whether the origin of the X-ray background had to 
be found in a diffuse hot IGM or in an unresolved population of discrete high-redshift 
sources. Since COBE, which did not find any trace of spectral distortions due to 
a hot IGM, the point source interpretation has quickly shown to be the 
proper one. ROSAT resolved $70\%$ of the soft XRB into point sources, many of 
which belong to a QSO population at $z \approx 1.5$, while others are probably 
obscured AGNs. The recent Chandra deep field results showing that more than $80\%$ 
of the XRB in the $2-10\hbox{\rm keV}$ is due to resolved sources nearly 
appear as ``famous last words'' on the matter. On the other hand, 
the fact that only 8 X-ray sources could be identified in the Hubble Deep Field 
of 3000 optical galaxies should be seen as the start of a new era, in which 
the X-ray observations will help us to gain extremely important insight into 
the formation of galaxies. Georgantopoulos described the Herschel Deep Field 
project which forms an attempt to get a better idea of the nature of the 
various X-ray emitting sources. QSOs and AGNs are obvious candidates, but also 
early-type galaxies may be avid X-ray emitting sources. Problematic 
identifications may concern X-ray emitting QSOs, whose redshift $z>5$ prevents 
them to be visible in optical images. In my view, such studies are extremely 
important in unravelling the nature of QSOs and AGNs, to find their differences 
as well as their similarities to normal galaxies. Still, we seem to have no 
understanding whatsoever of why such ``cosmic beasts'' get formed at all, that 
while we know that their formation will hold tremendous repercussions for their 
environment and for cosmic evolution in general. After all, their role in lifting the epoch 
of the ``Dark Ages'' may have been all-decisive and overriding !
\smallskip
\begin{figure}[t]
\centering\mbox{\hskip -0.truecm\psfig{figure=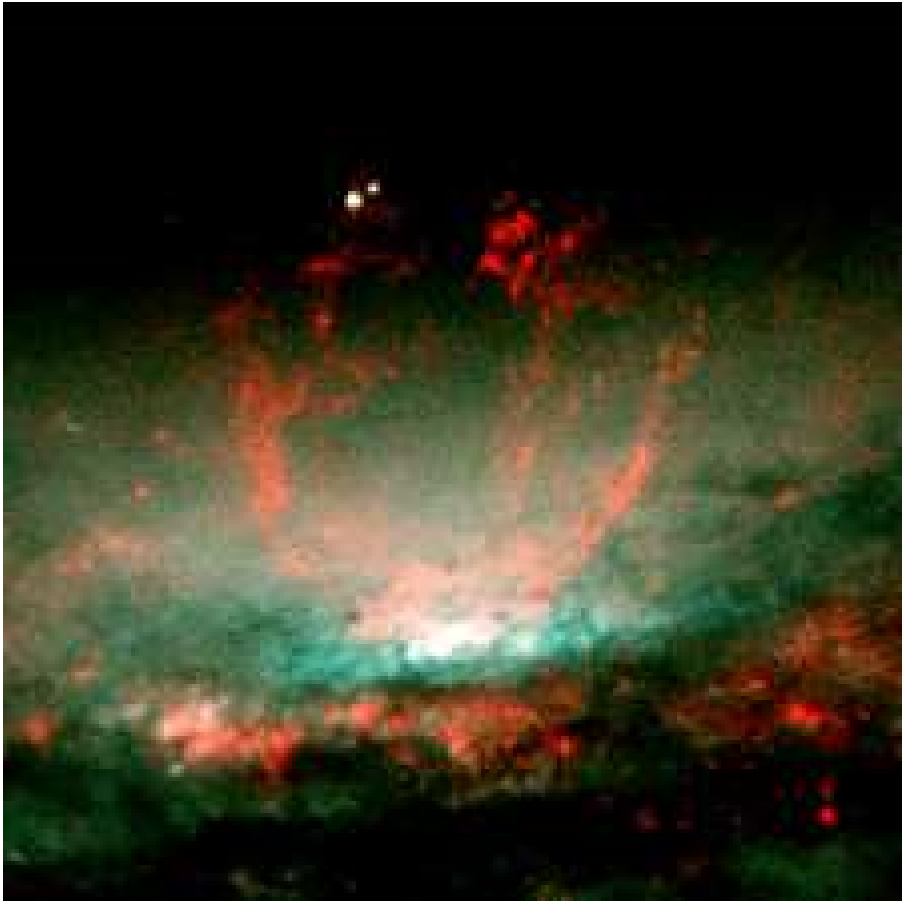,width=12.0cm}}
\vskip -0.25truecm
\caption{The gastrophysical Universe: dramatic events in the core of the galaxy NGC 3079, 
where a hot bubble of gas is rising from a cauldron of glowing matter. NGC 3079 may 
semble to a reasonable extent the dramatic phenomena taking place in young galaxies in the 
high-redshift Universe. Huge bursts of star formation, triggering violently 
driven winds, will be but one of the many diverse manifestations marking the 
Universe's youthful years. Credit: NASA, STScI, G. Cecil, 
S. Veilleux, J. Bland-Hawthorn, A. Filippenko.}
\end{figure}
\smallskip
\subsection*{{\it Infrastructure: \hskip 0.5cm Gas, Stars and (G)Astrophysics}}
The jewels in the `Crown of Creation' (Jefferson Airplane 1968). This obviously 
subjective description of the function of {\it galaxies} in the global context of the 
Cosmos is certainly not an entirely unjustifiable opinion. Galaxies certainly 
are amongst the most beautiful, complex and awe-inspiring creations in the Universe. 
They are largely self-organizing entities, cosmic cities harbouring a rich variety of 
ingredients, such as gas clouds in highly diverse physical states, stars and 
planets. All these aspects are elements of a highly complex state of 
organization, connected through an ingeneous and intricate ``life'' cycle of 
mutually influencing physical processes, phenomena and manifestations. 
Given the almost perfectly homogeneous Universe at recombination, it may 
be clear that the rise and growth of such complex entities in the Universe 
poses a daunting challenge for science. As ``clean'' gravitational 
studies have cleaned large parts of the road, succesfully breaching many 
hurdles, astrophysics is gradually preparing itself for the considerably more 
daunting world of ``gastrophysics''. 

In marked contrast to the situation with respect to studying the Megaparsec 
scale structure, where observational and theoretical advances went hand-in-hand, 
the complexities involved with galaxy formation processes are so enormous that 
the field is practically exclusively driven by observations. This may be no 
surprise, given the fact that the astrophysical processes involved with 
shaping the galaxies are so complex and of such wide variety. Gravity is no longer 
alone in beating the drum. Gas dynamical processes, radiative generation and 
transfer processes, the still even in our own Galactic circumstances largely 
ununderstood stellar formation processes, feedback by phenomenon such as supernovae 
and stellar winds, the chemical enrichment going along with the evolving stellar 
populations, its transport throughout a galaxy, the influence of magnetic fields 
are but a few of the examples of important contributors to the shaping of a galaxy. 

In a nice sequel of related contributions Van der Werf, Ivinson and Papadopoulos 
presented a broad and profound overview of the present-day status of our insight 
into the cosmic star formation history (Van der Werf), the kinematic evolution 
of galaxies (Van der Werf), the nature and identification of high redshift 
galaxies (Ivinson), and the ultimate source of cosmic beauty and shining (galaxies 
and stars alike), the nature and constitution of gas in the Early 
Universe (P. Papadopoulos). 

Concerning the basic constituents of the cosmic star formation history, 
the state of the gaseous component may arguably be seen as the most 
fundamental one. In particular the molecular phase is tightly coupled 
with the fueling of the subsequent star formation. P. Papadopoulos presented his 
instrumental work on the presence of molecules at high redshifts. 
From the nearby Galactic circumstances 
we know that gas has to get congregate into giant molecular clouds, 
which ``fragment'' into a whole range of compact ``subclumps''. In such 
dense, optically thick, circumstances become favorable for the ultimate 
collapse into a gaseous object that manages to generate its own energy 
by thermonuclear reactions: a star is born. However, we have but a 
superficial view of the situation in the outer reaches of the Universe. 
The basic molecular component of molecular clouds are evidently the simplest 
ones around, $H_2$. However, these suffer from notorious symmetry, 
so that we cannot exploit the radiation involved with ``dipole'' transitions. 
Luckily, CO may operate as a tracer molecule. The calibration of the 
tracer may still involve considerable uncertainties. Yet the detection of 
CO emission in 4C60.07 at $z=3.79$, probably involving a primeval merger, 
may certainly be considered a major achievement ! It has established 
observations of molecules as a fantastic way of studying the 
star formation circumstances in the high redshift Universe, out to 
redshifts of at least $z\approx4$ ! With ALMA getting into operation 
over the coming decade, a whole new era for the diagnostics of the 
physical conditions over a wide span of the cosmological timeline 
seems to have been opened up ! 

Turning to the very history of cosmological star formation itself, Van der 
Werf first pointed out that we do not even have a proper definition 
of what we mean by `` the formation of a galaxy''. A good working hypothesis 
may be to consider the onset of star formation in galaxies as such. It 
remains to be seen whether this is always a useful definition in case we have 
very early star formation in small clumps that only later fall into the larger 
peers that ultimately will start resembling the objects we nowadays describe 
as galaxy. Given the working definition, we may turn to the issue of star formation 
throughout the Universe. Van der Werf discussed the list of possibilities to trace such 
processes. These are based upon well-known symptoms of star formation in our 
immediate cosmic neighbourhood. Such tracers concern the emission of the 
newly formed stars themselves, the reprocessed radiation by the surrounding 
dust (SCUBA submm observations, COBE/DIRBE observations of the IR background) or 
the emission by surrounding gas excited or ionized by the newly formed star 
(H$\alpha$, L$\alpha$). With the help of corresponding observations at different 
redshifts one may (boldly) attempt to draw up a global cosmic star formation 
history. The pioneers investigations by Madau and Lilly have in the meantime been 
followed up by many related studies, incorporating the full arsenal of 
available observational information. 

Van der Werf described to some extent his own work on emission-line studies of 
cosmic star formation out to $z\approx 2.2$. This involved the 
first untargeted H$\alpha$ survey at $z=2.2$. Overall, his conclusion was 
that the Madau diagram is largely corroborated by most later work. Nonetheless, 
an important pitfall remains the fact that we do not really have an understanding 
of the physical circumstances surrounding star formation processes at high z. 
Theoretical understanding  will certainly have to play a crucial in converting 
the available circumstantial evidence into inescapable conclusions. The 
fact that we do not even seems to understand star formation locally remains 
a factor of suspicion. 

Extremely interesting was the evidence presented on the formation and (dynamical) 
evolution  of galaxy disks (Van der Werf). The question of the origin of galactic disks is 
evidently one of the most pressing within the whole of the galaxy formation 
process. Interestingly, we have learned that the large disks like that of our own 
Galaxy are certainly fully in place at rather moderate (cosmological) redshifts, 
$z\approx 0.5-1.0$. The intriguing work by N. Vogt, who managed to measure rotation curve of disk 
galaxies at $z \approx 1.00$, has indicated that even the Tully-Fisher relation seems 
to have settled by that epoch. Van der Werf described how one can follow up on these 
results by extending the study of the growth of disks to the higher redshifts at which 
it apparently must occur. He argued for H$\alpha$ as a perfect means of seeing the 
disk buildup at around $z\approx 2.$. Indeed a most thrilling prospect !

Ivinson subsequently gave a clear review of the various ways in which we can find galaxies 
at the even higher redshifts, beyond $z>2$. Being enabled to trace galaxies the hope 
is to infer at which epoch the galaxies formed their stars. Evidently, following the 
star formation history itself is an obvious first way. In the meantime, a variety of 
strategies have emerged as additional and complementary possibilities. Several colour 
and spectrum based techniques have materialized as effective selection methods 
for luminous galaxies. At radio wavelengths, one may for example look for sources 
with an ultra-steep spectrum. Over the past years the highly succesfull colour dropout techniques 
have evolved into a major industry for selecting out galaxies that are not only 
confined to the class of the very brightest galaxies. 

Particular emphasis was put on submm waveband observations. This benefits greatly from 
a welcome virtue of high-z dust emission, its negative k-correction. As a consequence the flux 
of such emission remains the same over a wide frequency range. The SCUBA instrument 
formed a major step in utilizing this possibility. A remaining problem is still 
the identification of sources in SCUBA observations with counterparts at other wavelengths. 
The identification with optical galaxies remains a cumbersome affair, as evidenced 
by the scarce overlap with the HDF image evidences. More promising appears to be 
the correspondence with radiogalaxies, leading to a prudent identification of 
SCUBA sources with highly clustered ellipticals at very high redshifts. They may 
therefore be th\'e way of probing the formation of galaxies in the very highest 
density regions, and hence of the formation of elliptical galaxies. 

Having arrived at the outer reaches of the observable yet familiar Universe, 
we had been offered a ``gastrophysical'' taste of the thrilling developments 
that are awaiting us on what must be one of the most challenging cosmic Odysseys 
in the history of astronomy. Also, we got to appreciate that the treasures which we 
may encounter on this voyage of discovery will be beautiful, rich, opulent.  
We may be justified in characterizing this as the search for the Holy Grail 
of cosmology and astrophysics. Perhaps, with some more nuance, the Holy Grail of 
Post-Planck Cosmology ... As yet we only obtained a very first impression of the 
first islands of the cosmic archipelago. Evidently, nothing has firmly settled yet, 
and this renders these explorations all the more exciting for the coming 
generations ... 

What a deception therefore to note the audacity with which P. Papadopoulos 
described it all as {\it ``Astro in a coffeecup ...''} This, with Greek 
gastronomy in mind shattered the high expectations of a mere barbarian from the 
north. A barbarian who for years had been indoctrinated with excessive laudations on 
the Greek cuisine ... yet, how fine it is then to have family who 
save the day and manage to repair such misunderstandings in preparing 
opulent tables filled with the most wonderful and delicious Epicurean delights ... 
\smallskip
\section*{\rm{\large CONCLUSION}}
Having arrived at the end of this wonderful and memorable cosmic Odyssey along 
the Cosmos, having been offered beautiful vistas along such richly 
varied directions, we maybe should 
properly finish with a return to the ancient masters who set us onto this 
never-ending voyage across space and time. A proper ``admonition'' 
to all scientists with pretensions too audacious, an advice to 
remain modest in the light of eternity, 
\bigskip
\vskip 2.0truecm
\begin{acknowledgments}
Looking back over the past months, the wonderful and enjoyable experience 
to which the organizers and these spring days of April 2002 treated us has 
grown into a happy and vivid memory of the best science has to offer. I feel 
truely priviliged to have attended a sparkling forum and `symposion'.

On behalf of all participants I would therefore like to express warm gratitude 
to the organizers, Manolis Plionis, Spiros Cotsakis and Ioannis Georgantopoulos. 
A gratitude which naturally extends to Koumentakou Ourania, whose support guaranteed 
the smooth and perfect organization. In this workshop, they provided all of us with 
a great example of how scientific curiosity and discussion can benefit from an 
environment of legendary Greek hospitality ... 
\end{acknowledgments}
\vfill\eject
\smallskip
\begin{figure}[t]
\centering\mbox{\hskip -0.truecm\psfig{figure=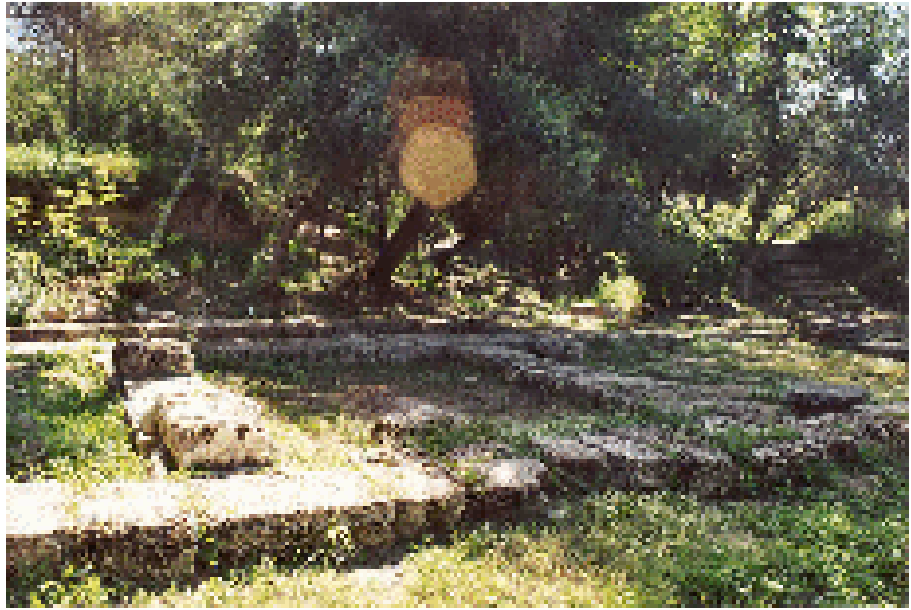,angle=270.,width=12.0cm}}
\vskip -0.25truecm
\caption{The Academia of Plato, anno 2001. ``The world of Becoming; everything in this world 
`comes to be and passes away, but never really is' '' (Plato, Timeaus). Yet, his world of 
ideas remained ever-Existent ! Notice how even nowadays the basic forms do show 
around the Academia ! photo: Rien van de Weygaert 2001}
\end{figure}
\vskip 1.0cm
\begin{flushright}
As for the world \\
-- call it that or cosmos or any other name acceptable to it --\\
we must ask about it the question one is bound to ask \\
to begin with about anything:\\
whether it has always existed and has not beginning, \\
or whether it has come into existence and started from some beginning.\\
The answer is that it has come into being;\\
for it is visible, tangible and corporeal,\\
and therefore perceptible to the senses,\\
and, as we saw,\\
sensible things are objects of opinion and sensation\\
$[...]$\\
Don't therefore be surprised, Socrates, \\
if on matters concerning the gods and the whole world of change\\
we are unable in every respect and on every occasion\\
to render consistent and accurate account.\\
\vskip 0.5cm
Plato (427-347 B.C.), {\it Timaeus}\\
\end{flushright} 
\vfill\eject                 
\begin{figure}
\vskip -0.25truecm
\centering\mbox{\hskip -0.truecm\psfig{figure=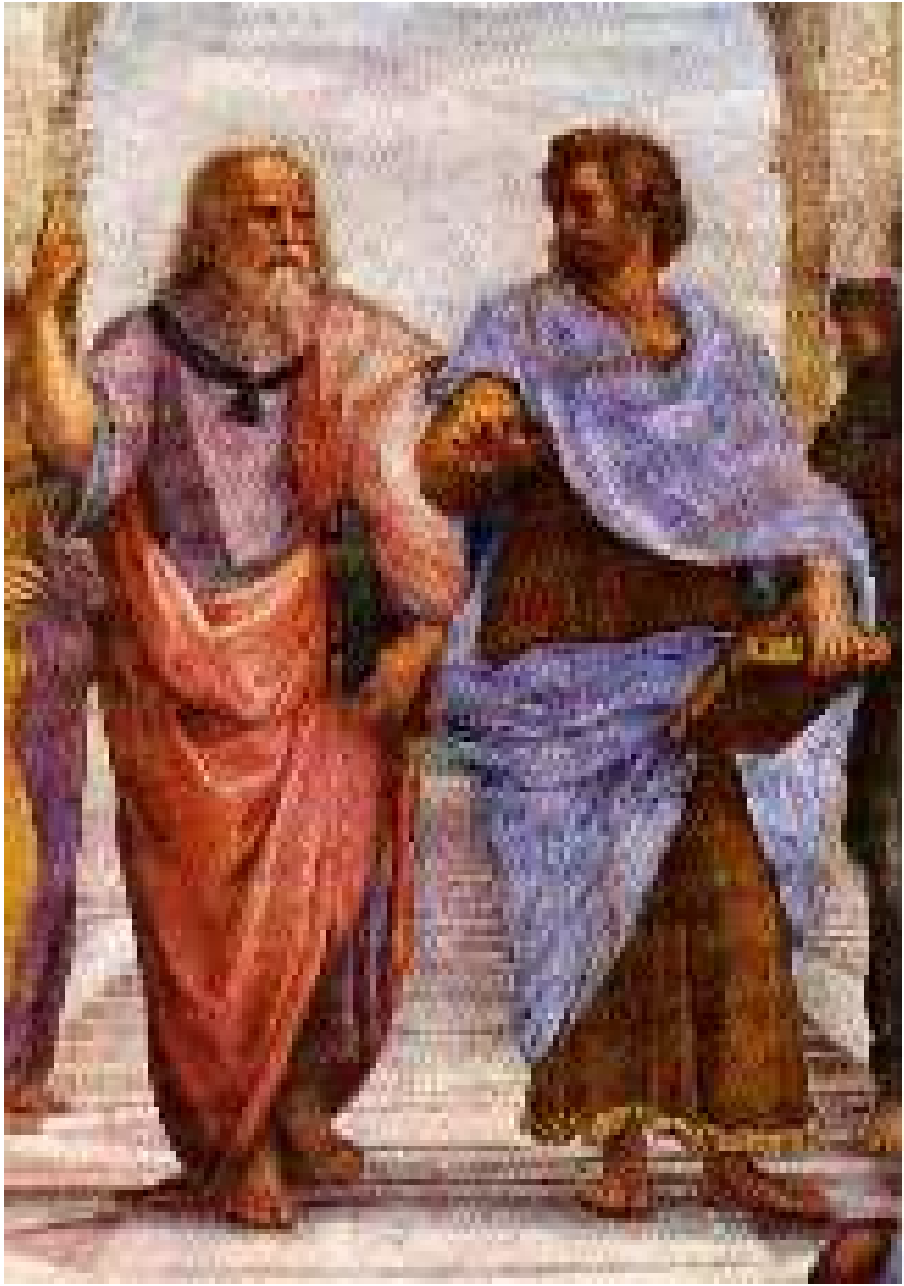,width=12.0cm}}
\vskip -0.25truecm
\caption{In Athens, Plato and Aristoteles conversing on the Universe. Detail from 
``Scuola di Atene'' in the Stanza della Segnature, Raffaello Sanzio (1483-1520). 
Courtesy: Christus Rex, Inc. and Michael Olteanu, MS.}
\end{figure}
\end{document}